\documentclass[12pt,preprint]{aastex}
\usepackage{epsfig}
\usepackage{graphicx}

\slugcomment{To appear in ApJL}

\shorttitle{Primordial Circumstellar Disks in Binary Systems}
\shortauthors{Cieza, L. et al.}

\begin{document}

\title{Primordial Circumstellar Disks in Binary Systems:  Evidence for Reduced Lifetimes}

\author{Lucas A. Cieza   \altaffilmark{1},
 Deborah L. Padgett\altaffilmark{2},
    Lori E. Allen \altaffilmark{3},
    Caer E. McCabe\altaffilmark{2},
Timothy  Y. Brooke\altaffilmark{2},
    Sean J. Carey\altaffilmark{2},
Nicholas L. Chapman\altaffilmark{4},
   Misato   Fukagawa\altaffilmark{5},
  Tracy L.  Huard\altaffilmark{6},
  Alberto   Noriga-Crespo\altaffilmark{2},
  Dawn  E.  Peterson\altaffilmark{3}, 
  Luisa M.  Rebull\altaffilmark{2},
}

\altaffiltext{1}{Institute for Astronomy, University of Hawaii at Manoa,  Honolulu, HI 96822. \emph{Spitzer} Fellow, lcieza@ifa,hawaii.edu}
\altaffiltext{2}{\emph{Spitzer} Science Center, MC 220-6, California Institute of Technology, Pasadena, CA 91125}
\altaffiltext{3}{Harvard-Smithsonian Center for Astrophysics, 60 Garden Street, MS 42, Cambridge, MA 02138}
\altaffiltext{4}{Jet Propulsion Laboratory, California Institute of Technology, 4800 Oak Grove Drive, MS 301-429, Pasadena, CA 91109, USA}
\altaffiltext{5}{Osaka University, 1-1 Machikaneyama, Toyonaka, Osaka 560-0043, Japan}
\altaffiltext{6}{Department of Astronomy, University of Maryland, College Park, MD 20742}

\begin{abstract}

We combine the results from  several  multiplicity surveys of pre-main-sequence stars 
located in four nearby star-forming regions with \emph{Spitzer}  data from three different 
Legacy Projects.
This allows us to construct a sample of  349 targets, including 125 binaries, 
which we use to  to investigate the effect of companions on the evolution of circumstellar disks. 
We find that the distribution of projected separations  of systems with \emph{Spitzer} excesses 
is significantly different (P $\sim$ 2.4e-5, according to the KS test for binaries with separations $<$ 400 AU)
from that of systems lacking evidence for a disk.  As expected,
systems with projected separations  $<$ 40 AU  are half as  likely to retain at least one disk than 
are systems with projected separations in the  40-400 AU range.
These results represent the first statistically significant  evidence for a correlation between binary 
separation and the presence of an inner disk (r $\sim$ 1 AU). 
Several factors (e.g., the incompleteness of the census of close binaries, the use of unresolved disk 
indicators, and projection effects)  have previously masked  this correlation in smaller samples. 
We discuss the implications of our findings for circumstellar disk lifetimes and the  formation of planets 
in multiple systems.

\end{abstract}

\section{Introduction}

Early multiplicity surveys of  pre-main-sequence (PMS)  stars  in nearby star-forming regions 
have established that most low-mass stars in the solar neighborhood form in multiple systems 
(e.g., Leinert et al. 1993; Simon et al. 1995). 
Understanding the effect of multiplicity on the evolution of  primordial circumstellar disks,
the birthplace of planets,  
is therefore crucial  to understand the potential for planet formation in most of 
the stars in the Galaxy.  
Recent models suggest that, starting  from a disk of planetary embryos and planetesimals,   
planets can form and survive around individual members of binary systems with separations 
as small as 5 AU (Quintana et al. 2007) as well as in orbits around both members of very 
close (r $<$ 0.3 AU)  binary systems (Quintana $\&$ Lissauer 2006). However, it is still  
unclear how the presence of a close companion affects the evolution of the accretion disk, 
and whether or not the early disruption of the primordial disk is the limiting factor for the formation 
of planets in multiple systems. 
  
Recent \emph{Spitzer}  studies (Padgett et al. 2006, Cieza et al. 2007) find that
up to 50$\%$ of  the youngest weak-line T Tauri stars (age $\sim$1 Myr) 
show  photospheric emission in the  mid-IR, which implies that their 
planet-forming regions are extremely depleted  of dust 
 at this early age.  One possible explanation for a very early disk  dissipation is the  
 efficient formation  of planets  in a $<$1 Myr timescale (e.g., Boss 2000).  
 Another possibility is the disruption of the disk due to the presence of 
close companions.  The outer disks around the  individual stars in a binary system are 
expected to be tidally truncated at a fraction ($\sim$0.3-0.5) of the binary separation 
(Papaloizou $\&$ Pringle 1977) and it has been suggested that the truncation of the 
disk limits the amount of material that can be accreted and effectively shortens the 
accretion timescale in close binaries (Ghez et al. 1993). 
 
Observational studies searching for a connection between close binaries and accelerated disk 
dissipation have so far yielded mixed and inconclusive results.
Ghez et al. (1993) performed a speckle imaging survey of 69 stars in Taurus and 
Ophiuchus and concluded that  the incidence of close binaries (r $<$ 50 AU) in 
non-accreting PMS stars  (i.e., weak-line T Tauri stars, WTTSs) is enhanced with respect 
to that of  accreting objets (i.e., classical T Tauri stars, CTTSs). However, subsequent studies 
have failed to confirm this conclusion. Leinert et al. (1993) and Kohler  $\&$ Leinert  (1998) performed 
larger speckle surveys  in Taurus and concluded that both the total  binary fraction and the  distribution of 
projected  binary separations of CTTSs and WTTSs are indistinguishable from 
each other. 
More recently, Ratzka et al. (2005) performed a speckle multiplicity 
survey of 158 PMS stars in Ophiuchus, the largest to date, 
and found a trend in the sense that objects with Spectral Energy Distributions (SEDs) 
suggesting the presence of a disk  have \emph{fewer} companions but at \emph{smaller}  
projected separations than that of objects with  SEDs consistent with bare stellar photospheres. 

Even though  the  truncation of disks in binary systems has very solid theoretical grounds, 
it has been repeatedly  argued that, as long as disks in multiple  systems can be replenished 
by an external reservoir (e.g., a circumbinary disk and/or an envelope), binaries will constrain 
the sizes of disks, but not necessarily reduce accretion disk lifetimes 
(Bouvier et al. 1997; Prato $\&$ Simon, 1997;  White $\&$ Ghez 1997, Mathieu et al. 1999).  
As a result, the lack of a statistically significant difference in the multiplicity and/or the 
binary separation of stars with and without disk indicators have lead many authors 
to conclude that multiplicity does not affect the lifetimes of  
circumstellar disks and/or planet formation (e.g., Simon $\&$ Prato 1995;  
Armitage et al. 2003;  Monin et al. 2007; Pascucci et al. 2008).  
Here we study a sample of  349 PMS stars, including 125 binaries with 
\emph{Spitzer} data, and show that  companions, at the peak of their separation 
distribution ($\sim$30 AU), do in fact shorten the lifetimes of circumstellar disks. 

\section{The Sample}

We collected projected binary separations from near-IR multiplicity surveys 
of PMS stars located in the following nearby (d $<$ 160 pc) star-forming regions:
Ophiuchus (Simon et al., 1995; Prato  2007;  Ratzka et al. 2005), Taurus 
(Leinert et al. 1993;  Simon et al. 1995; Kohler \& Leinert1998), 
Chameleon I (Lafreniere et al. 2008), and Corona Australis (Kohler et al. 2008). 

Most of the surveys were performed using  speckle imaging  (Leinert et al. 1993;  
Kohler \& Leinert 1998;  Ratzka et al. 2005;  Kohler et al. 2008),
but we also include results from lunar occultation  (Simon et al. 1995),  
radial velocity  (Prato  2007), and adaptive optics (Lafreniere et al. 2008) surveys. 
The speckle and adaptive optics surveys are sensitive to binaries down to projected 
separations  similar to the K-band diffraction limits of the  3.5-8.2 m telescope used 
 ($\sim$0.06-0.13$"$), while the lunar occultation observations 
 can detect binaries with separations as small as 0.005$"$. The radial velocity 
 surveys are of course sensitive to binaries at arbitrarily small separations, but are
 much less sensitive to wider binaries. 
As a result, our sample is highly heterogeneous  in terms of completeness. 
However, as it will discussed in Section 4., the detection biases of the different 
surveys included in our study are unlikely to affect our conclusions. 
 
Using their 2MASS coordinates and a 2$''$ matching radius, we searched for the  \emph{Spitzer} 
fluxes of all the targets from the surveys  discussed above in the catalogs produced by the  \emph{Cores to 
Disks} (Evans et al. 2003), \emph{Taurus} (Padget et al. 2006), and \emph{Gould Belt} (Allen et al. 2007) 
Legacy Projects. We focus on the IRAC 3.6 and 8.0 $\mu$m fluxes because (1) these \emph{Spitzer} Legacy surveys 
are sensitive enough to reach the stellar photospheres of  virtually all of the mutiplicity targets in all IRAC bands, 
but not so at the MIPS wavelengths, and (2) the [3.6]-[8.0] color  is the best disk indicator of all  IRAC colors. 
We found 3.6 and 8.0 $\mu$m fluxes with S/N $>$ 5 for 349 of the mutiplicity 
targets, including 125  binaries. Their  coordinates, projected separations 
(in the case of binaries), and \emph{Spitzer} fluxes are listed in Table 1.  

\section{Results}

\subsection{Disk Identification}

In order to investigate the effect of binaries in the lifetime of circumstellar disks, 
we first need to establish the presence or absence of a  circumstellar disk 
in each one of the systems in our sample.  We do so by using the \emph{Spitzer} colors
as a disk indicator, as shown in  Fig. 1.
There is a clear break in the color distribution of the sample around   [3.6]-[8.0]  = 0.8.
Thus, we consider systems  with  [3.6]-[8.0]  $<$ 0.8 to be  disk-less and  
systems with  [3.6]-[8.0]  $>$ 0.8  to harbor \emph{at least one} circumstellar disk.  

Given the distances involved (125-160 pc) and  \emph{Spitzer}'s limited resolutions 
(2.0$"$ FWHM at 8.0 $\mu$m), the vast majority of the  multiple 
systems remain unresolved. As a result, except for very wide separation 
systems,  \emph{Spitzer} provides no information on whether the IR 
excess originates from one or both of the components in a binary 
system.  
The dotted vertical line in Fig. 1 corresponds to 2.4$''$, the size
of 2 IRAC pixels, which is the radius of the photometry apertures for the 
the \emph{Taurus} Legacy Project data we use.  The lower S/N components 
of multiple objects detected within 2 pixels of each other have been dropped 
from their catalogs.  The \emph{Cores to Disks} and \emph{Gould Belt}   
teams performed PSF fitting photometry,  but objects less than 2 pixels apart  are still  unlikely to 
be resolved. 

From Fig. 1,  we find that 186 of the 349 objects listed in Table 1 have an IR-excess 
indicating the presence of a disk,  of which 72 are known to be binaries and 114 are apparently 
single stars.  Combining the multiplicity and disk identification information,  we find that  
the disk fraction of multiple stars is marginally \emph{larger} than 
that of  stars that appear to be single (57.6$\pm$4$\%$ vs 50.9 $\pm$3$\%$).
Taken at face value, this result seems to imply that multiplicity has no effect on the 
evolution of circumstellar disks. However, as it will be shown in the following sections, 
this initial result can easily be understood in terms of the incompleteness and biases 
of the multiplicity surveys and the limitations of the disk identification method.  

\subsection{The separation Distributions of Stars with and without a Disk}{\label{distri}}

The theoretical expectation of the effect of multiplicity on circumstellar 
disks is that, by tidally  truncating each others outer disks, close companions 
limit the amount of circumstellar material that can be accreted and
hence reduce the lifetimes of their disks (Papaloizou $\&$ Pringle 1977). 
Since the dispersal timescale of a  truncated disk is given by the viscous timescale at the truncation 
 radius,  one expects  the lifetimes of disks in binary systems to 
 be a strong function of the binary physical separation. This prediction
 can be tested by investigating the disk fraction as a function of binary projected 
 separation, or conversely,  the distributions of binary separation of systems with and 
 without a disk. Such distributions are shown in Fig. 2, where the measured 
 separations have been converted into projected physical separations (in AU)   
 using the following distances: 125 for Ophichus (Loinard et al. 2008), 130 pc for 
 Corona Australis (Casey et al. 1998),  140 pc for Taurus (Torres et al. 2007), and 
 160 pc for Chameleon I (Whittet et  al. 1997).  
 
 Fig. 2 clearly shows that targets without an excess tend to have companions
 at smaller separations than targets with an excess indicating the presence 
 of a disk. 
 The disk fraction of the systems with separations  less than 40 AU  is 38.2$\%$$\pm$6$\%$, 
 while the disk fraction of systems with separations in the 40-400 AU range is 77.8$\pm$7$\%$. 
 This difference in the disk fractions is  4.3-$\sigma$.
This is a robust result as a two-sided Kolmogorov--Smirnov (KS) test 
 shows that there is only a 2.4e-5 probability that the distributions 
 of binary separations of targets with and without a disks have been drawn 
 from the same parent population.   
Targets with projected separations $>$ 400 AU have been 
excluded from these calculations because they are likely to be 
resolved by \emph{Spitzer} and therefore require a  different statistical analysis
than the rest of the sample. 
 
 For binary systems with separations smaller than the  \emph{Spitzer} beam, the disk 
 fractions estimated above are not an accurate representation of the true disk fractions 
 of the individual components of the systems.
Assuming that each component of a binary  system has the same individual probability, 
DF$_{ind}$, of retaining a disk, the resulting fraction of systems with an IR excess, 
DF$_{sys}$, is given by the following equation: DF$_{sys}$=1--(1--DF$_{ind}$)$^2$.
Based on this formula, the disk fraction of the individual components of binaries
systems with projected separations $<$ 40 AU is 21.4$\pm$5$\%$, while that 
of  systems with separations in the  40-400 AU range is 52.9$\pm$7$\%$, a disk fraction
that is undistinguishable from that of apparently single stars, 50.9$\pm$3$\%$.
In reality, DF$_{ind}$ is unlikely to be exactly the same for both components of 
a binary system, especially if their mass ratio is high, but the above calculation 
illustrates well the limitations of the disk fractions derived from unresolved 
disk indicators.

\section{Discussion and Conclusions}

Although we interpret the strong correlation between disk  fraction and binary separation 
as an evidence for reduced disk lifetimes in close binary systems,  the census of binaries  
in our sample is incomplete, especially at small separations (see Fig. 2). 
Therefore, such a correlation could also arise if the older star-forming regions in our sample were 
observed with the techniques most sensitive to tight companions.  In that case,
the close binaries in our sample would be systematically older than the wide binaries and 
thus would  have lower disk fractions.  
We investigate this possibility by  examining the relative ages of the stars in the four different 
regions included in our study.  Instead of adopting ages  from the literature, which are known to 
be model dependent,   we derive their relative ages from their disk fractions. Table 2 shows 
that the targets from Cham I  and Ophuichus  have very similar disk fractions, 
but that the disk fraction  of Taurus  objects is clearly larger than that of CrA targets. 
Table 2 also shows that 1)  virtually all the lunar occultation and radial velocity  data,
which are the  most sensitive to close companions, come from the two youngest regions, and 2)
the  disk fraction of the lunar occultation plus radial velocity samples (LO+RV)  are almost 
identical to  that of the speckle plus adaptive optics samples (SP+AO). These two facts strongly 
suggest that our results are not affected by the  incompleteness and detection biases of our 
heterogeneous sample. 
We therefore conclude that the correlation between  disk fraction and binary separation 
is due to the effect close binaries have on primordial disk lifetimes.

\subsection{Implications for Disk Lifetimes}

It has already been shown that  $\lesssim$ 50-100 AU  separation binaries tend 
to have less (sub)millimeter emission than single stars or wider binaries 
(Osterloh $\&$ Beckwith 1995; Jensen et al.  1996;  Andrews $\&$ Williams, 2005).  
This implies lower disk masses for close binaries, but does not rule out the existence 
of small (r  $<$ 30 AU) disks with surface densities large enough to allow the formation 
of planets  (Mathieu et al. 2000).   Since \emph{Spitzer} IRAC data probe circumstellar 
distances of the order of 1 AU, our results show that close binaries not only reduce 
the sizes of  disks, but also their lifetimes. 

The fraction of stars with disks as a function of age observed in nearby 
star-forming regions shows that there is a very wide range of primordial 
disk lifetimes. Some stars lose their disks well within the first Myr, while others
retain their primordial disks for up to 10 Mys (Haisch et al. 2001; 
Cieza et al. 2007). 
The results from the previous section strongly suggest that
reduced disk lifetimes in binary systems  can account for a 
significant part of the observed overall dispersion in disk 
lifetimes.

The distribution of physical separations, $a$,  in solar-type PMS binaries is 
expected to peak  around 30 AU (Duquennoy $\&$ Mayor, 1991).   However, the disks 
around most binary systems will  have a truncation radius, R$_T$ given by 
R$_T$= 0.3-0.5$\times a  \sim$10-15 AU (Papaloizou \& Pringle, 1977).  These truncation 
radii are $\sim$10 times smaller than the typical  radii of disks around 
single stars (Andrews $\&$ Williams, 2007). 
The viscous timescales for a disk with a power-law surface density  profile of 
index p,  is given by t(r) $\propto$ r$^{2-p}$. Adopting p = 1, which is consistent
with both an steady state accretion disk and current observational
constraints (Andrews $\&$ Williams, 2007) leads to  t(r) $\propto$ r.
This implies that the lifetimes of disks around the individual
components of most binary systems  should be $\sim$10$\%$ of those
of single stars. 
Assuming disk lifetimes of 3-5 Myr for single stars, this 
corresponds to disk lifetimes of 0.3-0.5 Myr for binaries 
systems at the peak of their separation distribution. 
These short disk lifetimes  are  broadly consistent with the disk fraction 
of  $\sim$20$\%$ that we estimate for the individual components of binaries
systems with projected separations $<$ 40 AU (see Sec 3.2)

The rotation period distributions of PMS stars provide additional evidence  for very 
early disk dissipation in a significant fraction of them.  Rebull et al. (2004) and 
Cieza $\&$ Baliber (2007) both find that the bimodal period distribution of the Orion Nebula 
Cluster can only be reproduced, in the context of the disk regulation of angular momentum 
paradigm, if disk lifetimes are themselves bimodal, with 30-40$\%$ of the 
stars losing their regulating disks within $<$1 Myr of their formation.  
Such bimodal distribution is in fact expected from a population combining 
single and binary stars.  

\subsection{Implications for Planet Formation in Multiple Systems} 

Since most stars in the Galaxy are likely to form in multiple systems,
our results on the effect of multiplicity on circumstellar disk lifetimes have
direct implications for  planet formation.  It is now  almost universally accepted that 
terrestrial planets form through the continuous growth of solid particles.
However, there is much less of a consensus on the formation mechanism 
of  giant planets, with core accretion (Pollack et al. 1996) and gravitation 
instability (Boss 2000) being the two competing leading theories. 

Even though recent core accretion models  (e.g., Alibert et al. 2004) can reproduce  the formation of 
a solid core massive enough ($\sim$10 M$_{\oplus}$) to start accreting a gaseous 
envelope  within a few Myrs,  disk lifetimes of 0.3-0.5 Myrs are most likely to represent a 
significant  challenge to  giant planet formation through core accretion in binary 
systems.  
Reduced accretion lifetimes impose a weaker but still significant  obstacle to 
terrestrial planet  formation.  Gas drag is crucial for the growth of planetesimal 
into planetary embryos as it reduces the relative velocity of  planetesimals to a regime 
in which accretion can take place (Xie $\&$ Zhou,  2008). In the absence of gas,  
planetesimals  subjected  to the gravitational perturbations of  a binary system are expected 
to sustain relative velocities  well in excess of the threshold at which erosion 
dominates over accretion (Marzari $\&$ Scholl, 2000).  Thus, in order for terrestrial 
planets to efficiently form in binary systems, planetary embryos should be formed before the gas 
dissipates.

 While it seems very likely that multiplicity disfavors
 terrestrial planet formation in general  and giant planet formation through core accretion, 
 its effect on giant planet  formation through gravitational instability remains unclear.  
Depending on the details of the models, it has been argued both that multiplicity inhibits 
(e.g., Mayer et al. 2005) and enhances (e.g., Boss 2006) planet formation 
through gravitational instability.  

Given the limited number of currently  known exoplanets in binary systems, 
it is not yet possible to draw statistically significant  results on the incidence 
of planets as a function of binary separation. However,  there are already some 
indications that giant planets are in fact underrepresented in 
binary systems with separations $\lesssim$50-100 AU (Eggenberg et al.  2007; Bonavita $\&$ Desidera 2007). 
If most giant planets  form through core accretion, the findings 
presented  herein would provide a natural explanation for such a result.  

Even though many previous studies of smaller samples have searched for a connection between 
multiplicity and an accelerated disk dissipation (e.g.,   Leinert et al. 1993,  Ghez et al. 1993; 
Simon $\&$ Prato 1995;  Ratzka et al. 2005;  Bouwman et al. 2006; Monin et al.  2007), the results 
from this Letter  represent the first statistically  significant (i.e., over 3-$\sigma$) evidence that binaries 
reduce the lifetimes of circumstellar disks.  This is not too surprising considering that: 
(1) unresolved disk indicators 
only provide information on whether at least one of the components in a binary 
system has a disk, (2) the observed projected separations represent the \emph{minimum} 
possible value of the true  physical separations; thus many of the systems only appear to be 
tight binaries due to projection effects, 
and (3)  the multiplicity surveys are highly incomplete for tight systems, where the effect
on disk lifetimes is expected to be most severe. The combination of these three factors 
explains why a very large sample is necessary  to reveal the effect of multiplicity on disk 
lifetimes.

\acknowledgments
We thank Jonathan Williams,  Jonathan Swift, Michael Liu, and the anonymous referee for 
their valuable comments. Support for this work was provided by NASA through
the \emph{Spitzer} Fellowship Program under an award from Caltech.
This work makes use of  data obtained with the \emph{Spitzer} Space Telescope,
which is operated by JPL/Caltech,  under a contract with NASA.

\newpage

\begin{deluxetable}{lrcccccccccccc}
\rotate
\tabletypesize{\footnotesize}
\tablewidth{0pt}
\tablecaption{Multiplicity and \emph{Spitzer} data}
\tablehead{\colhead{Name}&\colhead{Ra}&\colhead{Dec}&\colhead{Reg}&\colhead{Sep}&\colhead{$Ref^a$}
&\colhead{F$_{3.6}$}&\colhead{ Error$_{3.6}$ }
&\colhead{F$_{4.5}$}&\colhead{ Error$_{4.5}$ }
&\colhead{F$_{5.8}$}&\colhead{ Error$_{5.8}$ }
&\colhead{F$_{8.0}$}&\colhead{Error$_{8.0}$} \\
\colhead{}&\colhead{(J2000)}&\colhead{(J2000)}&\colhead{}&\colhead{($''$)}&\colhead{}
&\colhead{(mJy)}&\colhead{ (mJy) }&\colhead{(mJy)}&\colhead{(mJy)}
&\colhead{(mJy)}&\colhead{ (mJy) }&\colhead{(mJy)}&\colhead{(mJy)} }
\startdata
     V1095 Tau  &   63.3090  &   28.3196  &     Tau  &  ------  &  1  &  1.08e+02  &  3.95e-01  &  7.17e+01  &  2.36e-01  &  4.96e+01  &  1.98e-01  &  2.73e+01  &  9.75e-02  \\
     V1096 Tau  &   63.3635  &   28.2736  &     Tau  &  ------  &  1  &  3.59e+02  &  6.58e-01  &  2.28e+02  &  5.54e-01  &  1.56e+02  &  5.87e-01  &  9.44e+01  &  1.82e-01  \\
     V773 Tau  &   63.5538  &   28.2034   &     Tau  &   0.170  &  1  &  1.10e+03  &  1.75e+00  &  1.03e+03  &  2.02e+00  &  1.02e+03  &  1.42e+00  &  1.13e+03  &  1.35e+00  \\
      FM Tau  &   63.5566  &   28.2137      &     Tau  &  ------  &  1  &  1.63e+02  &  3.71e-01  &  1.53e+02  &  3.74e-01  &  1.31e+02  &  6.16e-01  &  1.73e+02  &  4.10e-01  \\
     FN Tau  &   63.5608  &   28.4661        &     Tau  &  ------  &  1  &  2.59e+02  &  5.01e-01  &  2.43e+02  &  5.47e-01  &  2.37e+02  &  8.31e-01  &  3.22e+02  &  6.67e-01  \\
      CW Tau  &   63.5708  &   28.1827      &     Tau  &  ------  &  1  &  1.23e+03  &  1.86e+00  &  1.22e+03  &  2.55e+00  &  1.07e+03  &  1.60e+00  &  1.01e+03  &  1.15e+00  \\
         CX Tau  &   63.6994  &   26.8031     &     Tau  &  ------  &  1  &  1.14e+02  &  3.06e-01  &  1.01e+02  &  2.09e-01  &  9.76e+01  &  5.97e-01  &  1.42e+02  &  3.63e-01  \\
      V1098 Tau  &   63.6999  &   27.8763  &     Tau  &   0.470  &  1  &  3.43e+02  &  5.61e-01  &  2.09e+02  &  4.42e-01  &  1.42e+02  &  6.05e-01  &  8.21e+01  &  1.70e-01  \\
        FO Tau  &   63.7053  &   28.2085       &     Tau  &   0.165  &  1  &  2.73e+02  &  6.14e-01  &  2.44e+02  &  6.31e-01  &  2.35e+02  &  7.49e-01  &  2.74e+02  &  6.70e-01  \\
      V1068 Tau  &   64.1171  &   28.1266  &     Tau  &  ------  &  1  &  1.50e+02  &  3.95e-01  &  9.66e+01  &  2.20e-01  &  6.99e+01  &  2.40e-01  &  3.86e+01  &  1.21e-01  \\
\enddata
\tablecomments{The complete version of this table is in the electronic edition of the Journal. \\
$^a$References:  (1) Leinert et al.  1993;  (2) Simon et al. 1995;  (3) Lafreniere et al. 2008;  (4)  Prato et al. 2007;  (5) Ratzka et al. 2005;  (6)
Kohler et al. 2008.
}
\end{deluxetable}

\begin{deluxetable}{lccccccccclc}
\tabletypesize{\footnotesize}
\tablewidth{0pt}
\tablecaption{Properties of the Sample}
\tablehead{\colhead{Sample}&\multicolumn{4}{c}{\# of  Targets by Technique}&\colhead{Targets}&\colhead{Sin}&\colhead{Bin}&\colhead{Disk}&\colhead{Diskless}  &\colhead{DF}&\colhead{Relative Age}  \\
 \colhead{}&\colhead{SP}&\colhead{AO}&\colhead{LO}&\colhead{RV}&\colhead{(total \#)}&\colhead{(\#)}&\colhead{(\#)}&\colhead{(\#)}&\colhead{(\#)}&\colhead{(\%)}
 }
\startdata
    Tau       Region   &     65    &    0    &    46   &     0  &     84    &    43     &    41      &      54   &  30  &   64.3$\pm$5   & youngest    \\
    Oph      Region   &   141    &    0    &    33   &   19  &  158    & 108      &   50       &      83   &  75  &   52.5$\pm$4   &  young  \\
    Cham I Region   &     0      &   91   &     0    &     0   &   91     &   65      &    26       &     45  &  46  &   49.5$\pm$5    &  old  \\
    CrA      Region    &   16      &    0    &     0    &     1   &   16     &    8       &      8        &       4   &   12   &   25.0$\pm$11   & oldest\\
    \hline \\
    LO+RV sample  &     0        &    0   &    79  &   20  &        98  &     50    &    48      &     57  &   41   &   58.2$\pm$5 \\
    SP+AO sample  &   222     &   91  &      0  &      0 &       313 &   202    &  111     &   172  &  141  &  55.0$\pm$3 \\
\tablecomments{SP+AO+LO+RV  is  sometimes greater than the total number of targets because 
some overlap exists among some of the  sub-samples.} 
\enddata

\end{deluxetable}

\newpage

\begin{figure}
\figurenum{1}
\epsscale{1}
\plotone{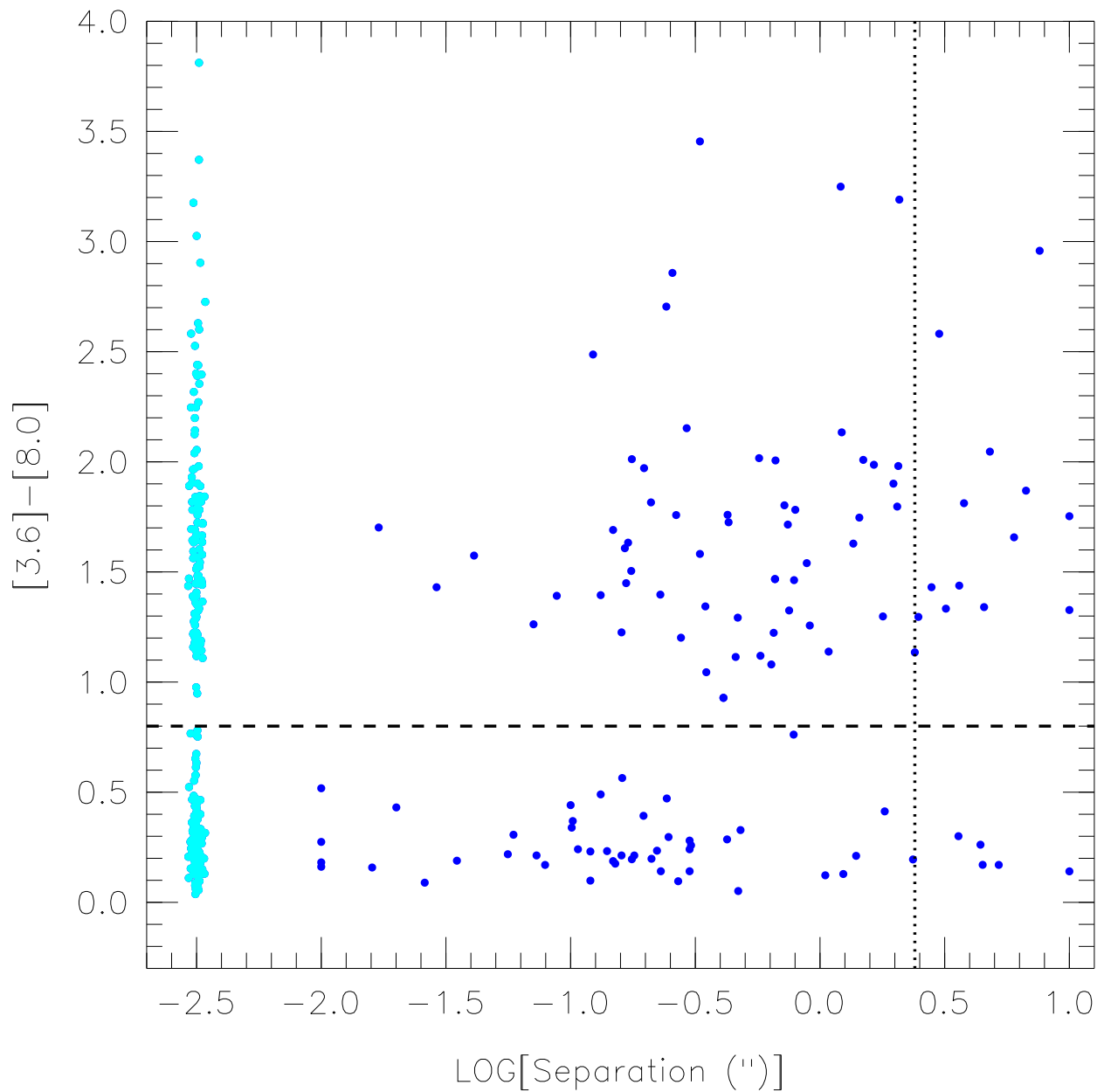}
\caption{
[3.6]-[8.0]  vs projected separation for our entire sample to illustrate our disk identification criterion.  
Systems with [3.6]-[8.0]  $<$ 0.8 are considered to be disk-less, while systems with  [3.6]-[8.0]  $>$ 0.8 
are considered to harbor \emph{at least} one disk. The dotted vertical line corresponds to 2.4$"$, the angle 
sustained by  2 IRAC pixels.   Only the few objects to the \emph{right} of this line are likely to be resolved 
by \emph{Spitzer}. Spectroscopic binaries have been  assigned a  separation of 0.01$"$.  Single stars, 
shown in light blue,  have been assigned a  logarithmic  separation of -2.5 plus very small random offsets 
to better show the density of objects at a given color.  
}
\end{figure}

\begin{figure}
\figurenum{2}
\epsscale{1}
\plotone{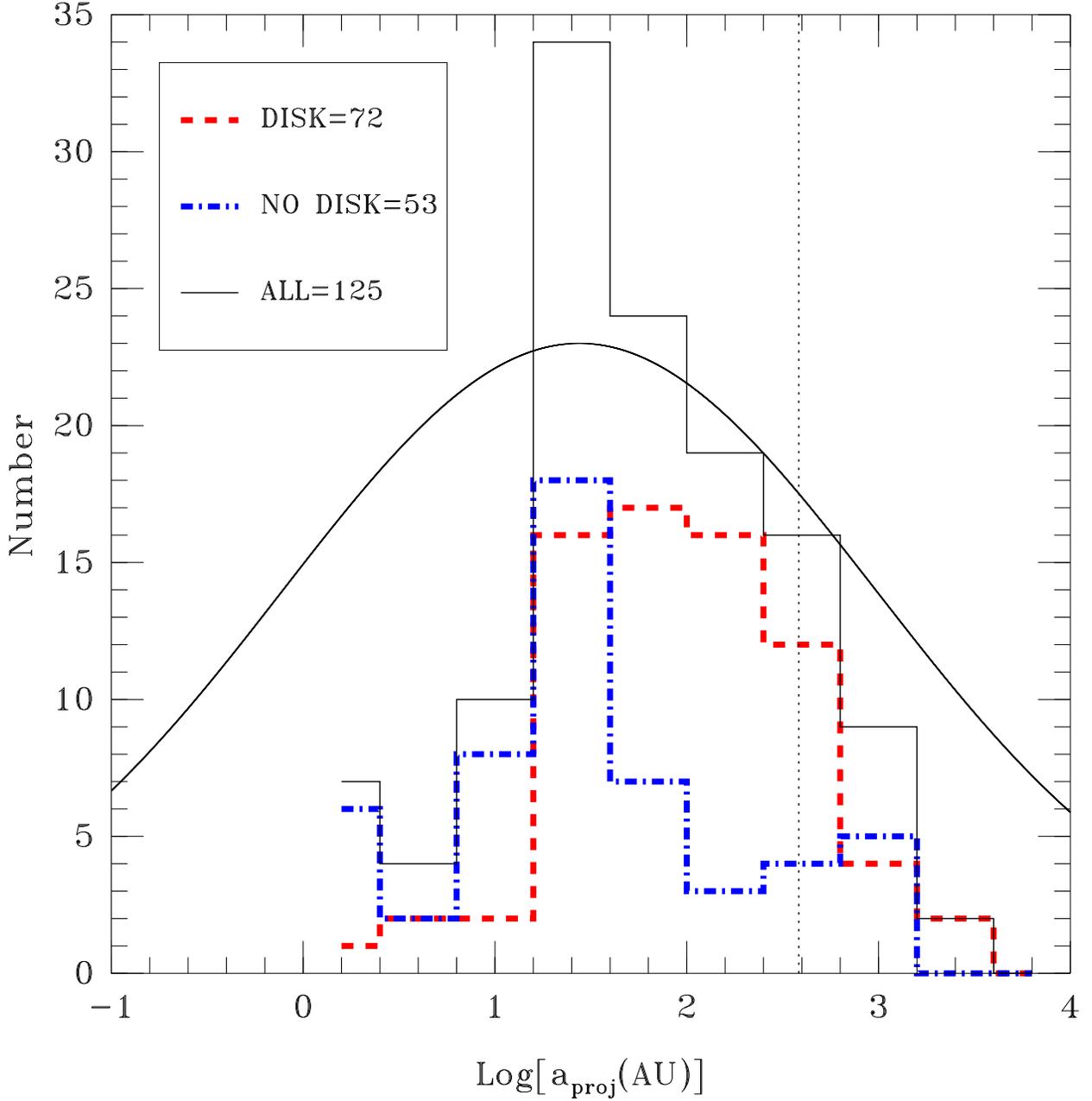}
\caption{
The histogram of projected separations for targets with and without  \emph{Spitzer} excesses indicating 
the presence of a disk.  Systems  without an excess clearly tend to have smaller separations.  
The solid curve  represents the distribution of binaries in solar-type stars (Duquennoy $\&$ Mayor, 1991). 
The census of companions is still highly incomplete for separations $\lesssim$ 20 AU.  
The vertical dotted line  at  X= 2.58 = LOG(384 AU)  corresponds to the resolution of 2 IRAC pixels (2.4$"$) 
at 160 pc, the distance  of the farthest regions in our sample.  Systems to the \emph{right} of this  line are likely to be 
resolved by \emph{Spizter}.  This explains their lower disk fraction, as we are measuring their 
DF$_{ind}$ instead of DF$_{sys}$.} 
\end{figure}

\end{document}